\documentstyle[12pt]{article}
\begin{document}

\title{Molecular Dynamics Simulation of Sonoluminescence: Modeling,
Algorithms and Simulation Results}

\author{Steven J. Ruuth\thanks{\mbox{Department of Mathematics, 
Simon Fraser University. ({\tt sruuth@sfu.ca}).}
          The work of this author was partially supported
          by DARPA and NSERC Canada.},~
Seth Putterman\thanks{\mbox{Department of Physics, 
University of California at Los Angeles. }
          The work of this author was partially supported by
	DARPA.}
 ~and Barry Merriman\thanks{\mbox{Department of Mathematics, 
University of California at Los Angeles. ({\tt barry@math.ucla.edu}).}
          The work of this author was partially supported
          by DARPA.}}
\pagestyle{headings}
\date{Dec. 30, 2000}
\maketitle

\begin{abstract}
Sonoluminescence is the phenomena of light emission from a collapsing
gas bubble in a liquid. Theoretical explanations of this extreme energy focusing are controversial
and difficult to validate experimentally.
We propose to use molecular dynamics simulations of the collapsing gas bubble to clarify the 
energy focusing mechanism, and provide insight into the mechanism of light emission.

In this paper, we model
the interior of a collapsing noble gas bubble as a hard sphere gas driven by
a spherical piston boundary moving according to the Rayleigh-Plesset equation.
We also include a simple treatment of ionization effects in the gas at high temperatures.
By using fast, tree-based algorithms, we can exactly follow the
dynamics of million particle systems during the collapse.
Our results clearly show strong energy focusing within the bubble, including the
formation of shocks, strong ionization, and temperatures in the range of 50,000---500,000 degrees Kelvin.
Our calculations show that the gas-liquid boundary interaction has
a strong effect on the internal gas dynamics. We also estimate the duration of the 
light pulse from our model, which predicts that it scales linearly with the ambient bubble radius.

As the number of particles in a physical
sonoluminescing bubble is within the foreseeable capability of molecular dynamics
simulations we also propose that fine scale sonoluminescence experiments can be viewed as excellent test problems for
advancing the art of molecular dynamics.
\end{abstract}

\section{Introduction}

\subsection{Background}
As a gas bubble in a liquid collapses, the potential energy stored during its prior expansion is released
and strongly focused. The extent of focusing can be so great that a burst of light
is emitted at the final stage of collapse. This process can be driven repeatedly by exciting bubbles
with a sound field, and the resulting transduction of sound into light is known as
sonoluminescence (SL) \cite{FrenzelSchultes,Temple,GaitanCrum,BarberPutterman,PuttermanWeninger,BarberHiller}. 

Sonoluminescence can be observed in dense fields of transient cavitation bubbles produced by
applying intense sound to a liquid, or in a periodic single bubble mode which allows more detailed
experimental observations.
In single bubble SL, a single gas bubble in the liquid is created and periodically
driven to expand and collapse by an applied sound field. The bubble begins its cycle of evolution
as the low pressure phase of the sound field arrives, causing it to expand to a maximal radius.
As the applied acoustic pressure increases, the bubble begins to collapse, first reaching its
ambient radius and corresponding ambient pressure and internal temperature, and then radially 
collapsing further, with the bubble walls falling inward driven by the rising external fluid pressure. The 
collapse accelerates rapidly, until gas trapped inside the bubble is compressed and heated to 
a pressure that ultimately  halts and reverses the motion of the bubble walls. Thus the bubble
reaches a minimum radius, and then rapidly ``bounces'' back to a much larger size. At some point
near the minimum radius, the resulting internal  ``hot spot'' 
can release a burst of light. While the basic bubble collapse dynamics can be observed for
a variety of gas and liquid combinations, light emission typically requires
the bubble to contain sufficient noble gas, and works particularly well in water.

The mechanism of light emission from the gas is not understood, nor is much known 
about related quantities such as peak temperatures, pressures or levels of ionization.

\subsection{Limitations of Hydrodynamic Models}
It is known that the mechanical conditions during collapse of a common gas bubble in water are quite extreme:
as the bubble reaches sub-micron diameter, the bubble wall experiences accelerations that exceed $10^{11}$g and 
supersonic changes in velocity that occur on picosecond time scales. 
In order to understand how this affects the state of the internal gas, the standard approach
is to apply continuum fluid mechanics. Some 
models assume that pressure and temperature are uniform inside of the 
collapsed bubble \cite{Hilgenfeldt} while other theories calculate the effects of
imploding shock waves \cite{BarberHiller,WuRoberts,Greenspan,Moss}. 
Various fluid models have been applied both at non-dissipative (Euler Equations)
and dissipative (Navier-Stokes Equations) levels of description \cite{KondicGersten,VuongSzeri,Yasui,MossYoung}. 

All these fluid approaches are limited in
their predictive power by the need to represent transport processes and the
equation of state. Under such extreme flow conditions, little is known about
these effects and one is forced to extrapolate from known forms. The net
result is that the modeling predictions directly reflect these assumptions.
This is not satisfactory
for the purpose of understanding what actually occurs within the bubble.

An especially fundamental limitation of continuum mechanical approaches
is the assumption of local thermodynamic equilibrium, i.e. it is assumed
that the macroscopic fluid variables do not change much over molecular
length and time scales.
Although the bubble starts out in such a state, its subsequent runaway 
collapse ultimately leads to a regime where this clearly does not hold. In this state, from which the ultraviolet 
picosecond flash of light is emitted, one can question the basic applicability of 
hydrodynamic models.

\subsection{Molecular Dynamics Modeling}
We propose to remove the assumption of thermodynamic equilibrium, and also eliminate any
controversy over the correct equation of state, by using molecular dynamics (MD) simulations
of the gas dynamics within the bubble. In this approach, we directly apply Newton's laws
of motion to the gas molecules, including as much detail as is desired (or practical) about the molecular
collisions and related atomic physics. While this approach is computationally intensive, it 
delivers to us a clear, physical picture of what actually transpires during the bubble collapse.
With the inclusion of sufficient detail and efficient programming, it could ultimately allow the simulation of 
of the light emission process itself.

While the small length and time scales of sonoluminescence present major obstacles
for hydrodynamic modeling, they actually make it ideal for molecular dynamics: precisely
because the final system is so small, it becomes possible to do a complete MD simulation of
the collapse. In fact, sonoluminescence is somewhat unique in this regard.
Usually the systems directly simulated with molecular dynamics 
are many orders of magnitude smaller---fewer particles, shorter time scales---than the 
corresponding systems realized in experiments or in nature, and this gap is too large to 
be eliminated by increases in computing power \cite{Rapaport}.
In contrast, the number of particles within the interior of a small SL bubble
is comparable to the number of simulation particles that can be handled with current computational facilities.  

For example, a typical SL bubble driven at 30kHz  has an ambient radius of 6 $\mu m$ and contains 
$2.25\times10^{10}$ particles. At the extreme, SL bubbles containing on the order of several million
particles have been observed in systems driven at Megahertz frequencies \cite{WeningerCamara}.
This compares well with simulations, where 
we have been able to compute the gas dynamics of a one million particle bubble collapse 
using a  run time of a few days on a single processor workstation-grade computer.
Parallel processing simulations would make 10 to 100 million particle simulations feasible.
As the number of simulation particles reaches that in real systems, the remaining 
computer power can be used to add in more complex atomic physics,
and thus allow more detailed study of the processes involved.

\subsection{Predictive Modeling Goals}
The overall goal of the MD modeling is to generate a better understanding of the 
processes that result in energy focusing and light emission during SL.
This is to be accomplished through a dual approach of model prediction and model validation:  
we use the model to illustrate the phenomena that cannot be experimentally observed during
the collapse, and also to make predictions that can be experimentally validated.

The basic experimental unknown in SL is the degree of energy focusing that is 
achieved inside of the bubble. For example, the spectral density of light from helium 
bubbles in water is still increasing at wavelengths as short as 200nm (energy exceeding 6 eV) 
where the extinction coefficient of water cuts off the  measurement \cite{BarberHiller}. 
Related to the question of energy focusing are the detailed questions of whether there is 
shock formation within the bubble, whether there is plasma formation, 
and what peak temperatures are achieved during the collapse. 
For example, the most extreme theoretical estimates suggest that
the interior may reach temperatures sufficient to induce deuterium-tritium fusion \cite{BarberWu}.
Over the range of parameter space studied, shock formation and strong ionization 
appear to be typical, and the {\em lowest} peak temperatures 
found in our simulations are about 40,000K, with the highest approaching 500,000K.
Our findings also indicate that boundary conditions strongly affect the 
interior motion. With a low, fixed temperature (i.e. heat bath) condition the peak 
temperatures and internal gradients are higher than for adiabatic motion. 

A key experimental observable in SL is the duration of the light flash, 
or ``flash width'' \cite{BarberPutterman}, because knowledge of this puts constraints
on the underlying mechanism of light emission. This can be used as a validation
point for any model or theory.
For example,  volume radiation
from a plasma will yield a different flash width than
surface radiation from a black body. Since our simulations
do not include  fundamental radiative mechanisms such as atomic excitation or charged particle acceleration, 
our current MD model cannot directly determine
determine the light emitting mechanism or the flash width.  However, a prediction about the 
flash width can be obtained from our calculation of the peak temperature as a function of time,
assuming the light emission occurs while the peak temperature is high.
Our simulations for Helium 
show that simple adiabatic compression does not produce a sharp temperature spike in time, 
but the thermal boundary condition causes a spike with a duration 
that scales with the ambient bubble radius.  This result predicts that the flash width should
scale with the ambient bubble radius.  If valid, this scaling suggests that at high acoustic
frequencies $[\sim 10$MHz] \cite{WeningerCamara} the duration of a SL flash could be about equal to or
less than 1 ps.

\subsection{Outline of the Paper}
The outline of the paper follows:  Section~2 describes the model for
the bubble collapse in detail.
Section~3 outlines the principle algorithms used to 
evolve the hard sphere system.
Section~4 provides detailed results 
from our MD simulations.  Finally, Section~5 concludes
a summary of our observations, and lists interesting 
future areas of investigation suggested by this first attempt 
at molecular dynamics modeling of sonoluminescence.

\section{Modeling Sonoluminescence Bubbles}
In this section, we present our Molecular Dynamics model 
for SL bubbles. The overall strategy is to model the system as a 
spherical piston that compresses a gas of hard spheres, with energy deducted 
from the system for ionization events at higher temperatures.
The details and motivations for this are given in the following subsections.

\subsection{Model Parameters}
We want to focus on the simulation of single bubble sonoluminescence, so
that results can be compared to the best studied experimental SL systems. Such
bubbles remain spherical during their collapse \cite{WeningerEvans}, and their behavior is
parameterized by their ambient radius (the radius they have when at rest at the
ambient pressure) and their maximum radius (the radius they attain when maximally
expanded at the low pressure point of the applied sound field).

We cannot directly simulate all such SL bubbles, since they may contain several orders
of magnitude more gas particles than our computational budget can accommodate.
Typically, we can afford to do a calculation
with some given number of simulation particles, $N$, and the question becomes how large of a bubble 
can we directly simulate.
The ambient radius, $R_0$, is related to the number of gas particles, $N$,
by the ideal gas equation of state
\[
P_0 \left( \frac{4}{3} \pi R_0^3 \right) = k T_0 N
\]
where $T_0 = 300K$ and $P_0 = 1$ atm are the ambient temperature and pressure,
and $k$ is Boltzmann's constant.
Thus we see that the fewer the simulation particles we use, the smaller the ambient
size of the bubble being simulated.

Once the ambient size is determined by our simulation budget, we are however free
to choose any maximum radius. For experimentally relevant simulations, 
the maximum radius $R_m$ is chosen to
yield the same ratio of $R_m/R_0$ [$\sim10$] for the MD simulation as is seen in experimental SL bubbles. 
This is natural because this ratio is a measure of the available energy stored in the expansion, since the stored 
energy/particle 
due to the work done by expanding to the maximum volume $V_m$ against the applied pressure $P_0$ 
is
\[
\frac{P_0 V_m}{N} = k T_0 \frac{R_m^3}{R_0^3}.
\]

\subsection{Bubble Collapse}
Since the bubble remains spherical during collapse, its boundary dynamics are described entirely by 
the radius as a function of time, $R(t)$. 
We are concerned with energy focusing processes and gas dynamics inside
the bubble, and in this spirit we will take $R(t)$ as being known. A
convenient model of the spherical piston that captures some qualitative
features of the supersonic collapse is provided by Rayleigh's equation \cite{PuttermanWeninger}
\begin{equation}
R\ddot{R} + \frac{3}{2} \dot{R}^2 = \left[ P_g(R) - P_0\right] / \rho, \label{eq:rayleigh}
\end{equation}
with a van der Waals hard core equation of state
\begin{equation}
P_g(R) = \frac{P_0 R_0^{3\gamma}}{(R^3 - a^3)^\gamma}, \label{eq:state}
\end{equation}
$\gamma=5/3$, where $a$ is the radius of the gas in the bubble when compressed to its van der 
Waals hard core ($R_0/a = 10.1, 9.15, 7.84$ for He, Ar, Xe), $\rho$  is the density of 
the surrounding fluid, and the initial condition for the solution to
(\ref{eq:rayleigh}) is that $\dot{R}=0$ when $R=R_m$.  We emphasize that 
the derivation of Equations 
(\ref{eq:rayleigh},\ref{eq:state}) applies only for small Mach number motion and thus they are 
invalid as a fundamental theory for SL \cite{Putterman}. 
However, in this first attempt to simulate SL with
molecular dynamics we are interested in possible focusing processes within
the bubble and use of (\ref{eq:rayleigh},\ref{eq:state}) as a launch condition appears appropriate since
the resulting $R(t)$ reasonably approximates the gross bubble pulsation \cite{PuttermanWeninger}.

Consistent with this approximation, viscous damping and
acoustic radiation have also been neglected.
At the next level of simulation one should include a self-consistently
determined boundary condition on pressure at the bubble's wall.
In this way, energy loss due to acoustic radiation is properly accounted for.

\begin{figure}
 \vspace*{2.in}
 \includegraphics{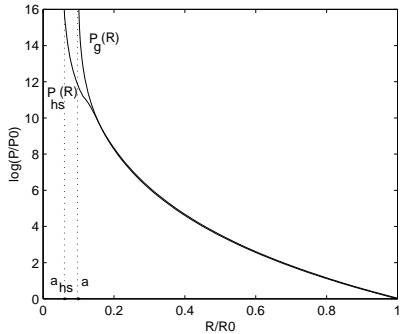}
 \caption{Plots of the adiabatic equations of state for the van der Waals pressure $P_g(R)$ and
the computed hard sphere pressure $P_{hs}(R)$ for Helium.} \label{fig:pg}
\end{figure}

As a point of comparison, it is worth noting that the adiabatic equations of state for the van der Waals pressure
$P_g(R)$ and the computed {\em equilibrium} hard sphere pressure $P_{hs}(R)$ (the equilibrium pressure as a
function of radius as the radius is decreased {\em slowly} on the hard spheres) agree very well, except
at bubble radii near the hard core, as graphed in Figure~\ref{fig:pg} 
for Helium (with the other noble gases also in good agreement, except
near the hard core).
At small radii, the van der Waals pressure diverges as $R$ tends to $a$ and
the $P_{hs}(R)$ diverges as $R$ tends to
\[
a_{hs} \equiv \left( \frac{N}{0.63}\right)^{1/3} \frac{\sigma}{2}
\]
which is the minimum radius for random closed packing \cite{RusselSaville}.

\subsection{Gas Dynamics} \label{sec:dynamics}
It has been observed that for SL in water, the 
bubble must contain sufficient amounts of a noble gas.
Thus in many single bubble SL experiments, the water is first de-gassed to
remove atmospheric gases, 
and then re-saturated with a noble gas to produce pure noble
gas bubbles. We will focus our gas dynamic model on this system, 
since it is a frequent experimental model and also because it
allows the simplest molecular gas dynamics models. Because the 
gas is noble, it consists of isolated atoms that do not engage
in chemical reactions. Thus we can model it with simple 
gas particles that have no rotational or internal vibrational
degrees of freedom, and which do not engage in any chemical reactions
with the water walls of the bubble, even at elevated temperatures.

Molecular dynamics simulations for such simple gas particles
fall into two broad categories,  defined by the way they treat
interatomic forces. The forces can either be given by a potential
that varies continuously with radius from the atom center (``soft sphere''), or
by a potential that is a step function of radius (``hard sphere''). The latter particles
behave simply like billiard balls. While the continuous
potential are more physically realistic, they are also much more costly 
to compute with.
This is because numerical time integration methods must be used to 
compute the particle motions in response to the continuously varying forces,
and the time step  must be small enough to
accurately resolve all particle trajectories in the system. Thus
the motion of a few fast moving particles
will force the use a small, costly timestep for all
particles in the system.  In contrast, step potentials
do not experience this problem because they evolve in time
by a series of discrete collision events. No explicit numerical
integration is needed since impulsive collisions are carried out
only when atoms interact, and between collisions each atom
follows an independent linear trajectory. Thus each atom effectively
uses its own optimally large timestep, instead of an excessively
small step imposed by the the fastest particles in the system.  Moreover, there is
no numerical integration error because trajectories
are evaluated to within the roundoff error of the machine. \cite{Rapaport}

Because of this difference in computational cost, it is desirable to use the
hard sphere model if it can capture the physics of interest. In our case, we
want to get accurate gas dynamics at mid to high energies for fairly large numbers of
particles. Whether hard spheres are sufficient to model this regime in an SL bubble is an
empirical question, but such models have been shown to yield accurate predictions
of noble gas viscosity from room temperature up to the gas ionization temperatures \cite{BirdBook}.
We take this to be a reasonable validation that a hard sphere gas provides a good model
for the gas dynamics encountered during bubble collapse, at least up to 
ionization temperatures. Near that point and beyond, it also seems reasonable that 
a hard sphere model applies, since the softer parts of the potential are all the more 
insignificant for high energy collisions.

The dynamics of a hard sphere system involve processing
impulsive collisions at the collision times.
To illustrate, consider two particles separated by a relative position {\bf r}
and having a relative velocity {\bf v}.   These particles collide
if their separation equals the atomic diameter $\sigma$ at
some time $\tau$ in the future.  If such a collision occurs,
then $\tau$ is the smaller positive solution of
\[
\left|{\bf r} + {\bf v} \tau\right| = \sigma
\]
which has a solution
\[
\tau = -\frac{1}{{\bf v}\cdot{\bf v}} \left( {\bf r} \cdot {\bf v} + \sqrt{ ({\bf r}\cdot {\bf v})^2 - 
{\bf v}\cdot{\bf v}({\bf r}\cdot{\bf r}-\sigma^2)}\right)
\]
Collisions are carried out impulsively so that the change in velocities
preserves energy and momentum.  Specifically,
\[
\Delta {\bf v}_1 = - \Delta {\bf v}_2 = - \frac{({\bf r}_c\cdot {\bf v}) {\bf r}_c}{ \sigma^2}
\]
where $\Delta {\bf v}_1$ is the change in velocity of the first particle,
$\Delta {\bf v}_2$ is the change in velocity of the second particle
and ${\bf r}_c$ is the relative position at the time of collision.

Extensions to step potentials that consist of a hard repulsive
core surrounded by an attractive well are also possible.  See
\cite{Rapaport} for details.

\subsection{Bubble Wall Boundary Conditions}
When a gas particle hits the bubble wall, it might simply be directed back into the interior
by a strong collision with a liquid molecule, or it may penetrate into the liquid, undergoing
multiple thermalizing collisions. In the latter case, assuming the liquid is already saturated with
gas atoms, the thermalized atom (or an equivalent one from the saturated liquid reservoir) will 
ultimately random walk its way back into the bubble interior. 

For our MD model, we will idealize these two modes of boundary interaction as either
as energy conserving {\it specular} collisions or as a {\it heat bath} boundary conditions. 

For the case of specular collisions, particles
reflect from the boundary with a speed
equal to the collision speed in the local rest frame
of the wall.  The direction of propagation is determined
according to the law of reflection, where 
the angle of incidence equals the angle of reflection with respect to the local normal to the
bubble surface.

For heat bath boundary conditions, when a particle hits the boundary
it is assigned a thermal velocity at the ambient liquid temperature $T_0$,
and the direction of propagation back into the interior is chosen 
according to a suitable angular distribution.
We ignore the small time
lag that might exist between exit and reentry for the thermalized gas particle.

For the angular distribution, we use the {\it cosine distribution}, 
where the angle of reflection $\theta$ is assigned
randomly according to a probability density function,
\[
f(\theta) = \left\{ 
\begin{array}{ll}
(1/2)\cos \theta & \mbox{for $-\pi/2 \le \theta \le \pi/2$} \\
0 & \mbox{otherwise}
\end{array} \right.
\]
We have also tried a uniform distribution in angle in selected test cases.  This did not
significantly change the simulation results.  (Nonetheless, we note that there may exist
situations where results differ qualitatively
since a uniform distribution has a greater tendency to cause reflected particles to build up near the wall.)

In reality, we expect that the physical boundary
will have some characteristics of both models. By investigating these extreme cases 
we hope to see the full range of effects that boundary conditions can have on the bubble dynamics.

\subsection{Initialization}
Initially the bubble is its maximum radius, $(R=R_m)$, and particles are moving in uniformly
distributed random directions, with the same thermal speed $v_{th}(T_i)=\sqrt{3T_i k/m}$, where $T_i$ is the
initial temperature and $m$ is the mass of the particle.
Randomization of speeds is not necessary,
since the particles rapidly thermalize their energies in any case. For heat
bath boundary conditions, the initial temperature is taken to be
the ambient temperature, $T_i = T_0$, reflecting the thermalization with the liquid.  
With specular boundary conditions
this choice generates unphysically large temperatures
when the bubble collapses to its ambient size, due to adiabatic heating. In order to achieve the
ambient temperature $T_0$ at the ambient radius $R_0$,
the initial temperature must be scaled down to $T_i = T_0/(R_m/R_0)^2 = T_0/100$.
The factor of $(R_0/R_m)^2$ approximately cancels the adiabatic heating (since
$TR^2 = \mbox{constant}$ in a $\gamma = 5/3$ ideal gas at constant entropy) 
during the initial, slow portion of the collapse.

\subsection{Hard Sphere Properties}
The basic properties associated with the hard sphere model are the gas particle
mass and diameter. The mass is simply taken to be the mass of the noble gas atom being
simulated.  See Table~I below.  The choice of proper hard sphere diameter is a much more difficult question.
The diameter should represent the statistical
average distance of approach of the particles
during collisions, and thus in general it should depend on
the collision energy. 

In our most basic model we will neglect this temperature dependence
and choose particle diameters that have been derived from
the kinetic theory for the viscosity of
a gas at room temperature \cite{Jeans,Rankine1,Rankine2}:
\begin{center}
\begin{tabular}{|c|r|c|} \hline
Gas & Mass [g/mole] & Diameter [\AA] \\  \hline
He &   4.00 & 2.18\\
Ar &  39.95 & 3.66\\
Xe & 131.29 & 4.92\\ \hline
\end{tabular}
\vskip 0.1in
Table I. Hard Sphere Diameters and Masses
\end{center}

To produce a more realistic model for higher temperature regimes of interest in
SL, the hard sphere
diameter should depend on the relative velocity of
the colliding particles. A variety of models have been proposed to take this effect into account \cite{BirdBook}.  
These include the variable hard sphere (VHS) model \cite{bird},
the variable soft sphere (VSS) model \cite{Koura1,Koura2} and the generalized hard sphere
(GHS) model \cite{Hassan} which is an extension of the VHS and VSS models.
In this report, we are mainly interested in contrasting how variable and
constant hard sphere diameters affect our simulations, so the recent VSS model
is chosen for its combination of simplicity and calibrated accuracy.
In fact, we find that the VSS model and constant diameter
models often produce quantitatively similar results.  See Section~\ref{sec:results} for details.

The viscosity based diameter of a VSS particle is 
\begin{equation}
\sigma = \left( \frac{5 (\alpha+1)(\alpha+2)(m/\pi)^{1/2}(k T_{ref})^{\omega}}{16 \alpha \Gamma(9/2-\omega)\mu_{ref}
E_t^{\omega-1/2}}\right)^{1/2}
\end{equation}
where $k$ is Boltzmann's constant, $m$ is the mass of
the particle, $\omega$ is the dimensionless viscosity index and $\alpha$
is a dimensionless constant for each gas.  The constant $\mu_{ref}$ represents
the viscosity at the reference temperature $(T_{ref}=273K)$ and pressure (1 atm). Finally, $E_t=(1/2) m_r c_r^2$
is the asymptotic kinetic energy where $m_r$ is the reduced
mass
\[
m_r = \frac{m_{particle_1} m_{particle_2}}{m_{particle_1}+m_{particle_2}}
\]
and $c_r$ is the relative velocity between the particles.  Tabulated values
for these new parameters are provided in \cite{BirdBook,ChapmanCowling} and are summarized below.
\begin{center}
\begin{tabular}{|c|c|c|c|} \hline
Gas & $\omega$ & $\mu_{ref}$ [Nsm$^{-2}$] & $\alpha$ \\  \hline
He & 0.67 & $1.865\times 10^{-5}$ & 1.26 \\
Ar & 0.81 & $2.117\times 10^{-5}$ & 1.40 \\
Xe & 0.85 & $2.107\times 10^{-5}$ & 1.44 \\ \hline
\end{tabular}
\vskip 0.1in
Table II. VSS Molecular Parameters
\end{center}

\subsection{Ionization Effects}
Near the minimum radius of the bubble, collisions may become sufficiently
energetic to ionize the gas atoms. Ionization exerts a very strong cooling
effect on the gas, since on the order of $10$eV of thermal energy is removed
from the gas by each ionization event. Indeed, if such energy losses are
not included, Xenon simulations can
reach temperatures in excess of one million degrees Kelvin, while
the inclusion of ionization cooling brings these peak temperatures down substantially 
(see Section~\ref{sec:results}).
This clearly shows that some degree of ionization must occur during collapse, and that its
cooling effects must be included for proper prediction of peak temperatures.
The ions and free electrons produced by ionization will move according
to coulomb forces, but the need to incorporate these effects is not as
clear, and their inclusion is more difficult and expensive due to the long range effects, 
so they will not be included in this
first treatment. We will only consider the impact of ionization on energy accounting. 

For the purpose of energy accounting, an ionization ultimately produces two losses: the
energy of ionization is lost immediately, and the emitted cold electron will quickly
be heated to thermal equilibrium with the gas through subsequent electron-gas collisions, thus extracting
an additional one particle's worth of thermal energy by the equipartition of energy.

For our model, we will simply assume ionization
occurs with probability 1 whenever the collision energy exceeds
the ionization potential, we deduct a suitable amount of energy from the pair.
We also we keep track of how many electrons each particle has lost, so that we can 
make use of the appropriate next ionization energies and calculate the local ionization levels.
The direction of gas particle propagation is updated
exactly as without ionization.  See Section~\ref{sec:dynamics} for details.

More precisely, if the 
kinetic energy (in the center of mass frame) of two colliding particles is greater than the
next ionization energy of either of the pair (which may already be ionized), that particle
loses an additional electron.
We account for the net energy loss by 
setting the kinetic energy of the pair to be
\[
\frac{2}{3} \left( E_0 - \chi \frac{E_0}{E} \right)
\]
where $E_0$ is the original kinetic energy of the particle,
$E$ is the kinetic energy of both particles before the collision
and $\chi$ is the ionization potential of the minimally charged particle.  

Note that the final kinetic energy of the pair is the initial energy, minus the 
ionization energy, with an additional 1/3 deducted to represent the subsequent
energy lost to thermalizing the electron. This is not the only possible
way to include this effect, and of course in reality this process
involves losses from other gas particles besides the colliding pair, but this approach 
is the simplest way to include the effect.

We also do not account for
subsequent electron-ion recombination to neutral atoms, although this would be interesting to 
include at the next level of description. In particular, this could be an interesting source of 
radiation as the hot spot decays.

For reference, the approximate ionization potentials used for the three noble gases
are provided in Table~III below. 
\begin{center}
\begin{tabular}{|l|c|c|c|c|c|c|c|c|} \hline
Gas &  \multicolumn{8}{c|}{Ion} \\ \cline{2-9}
& Neutral & 1+ & 2+ & 3+ & 4+ & 5+ & 6+ & 7+  \\  \hline
He \cite{Handbook} & 2.37 & 5.25 &  &  &  &  &  &   \\  
Ar \cite{Handbook} & 2.08 & 2.67 & 3.93 & 5.77 & 7.24 & 8.78 & 12.0 & 13.8  \\  
Xe \cite{Handbook,Carlson} & 1.17 & 2.05 & 3.10 & 4.60 & 5.76 & 6.93 & 9.46 & 10.8 \\ \hline
\end{tabular}
\vskip 0.1in
Table III. Ionization potential [MJ/mol].  Each entry represents the 
energy required to ionize the indicated state.
\end{center}

\section{Algorithm}
Efficient algorithms are needed to evolve our hard sphere model for sonoluminescence 
since a naive coding is prohibitively slow for anything more than a few thousand particles.
To achieve this goal, we modify and extend existing methods \cite{Rapaport}
rather than develop new algorithms and codes from scratch.  
This section outlines the principle algorithms used to 
evolve our hard sphere system.
Further details and basic codes are provided in \cite{Rapaport}.

\subsection{Cell Subdivision}
The hard sphere simulation proceeds according to a
time ordered sequence of collision events \cite{Rapaport,erp77}.
But clearly a direct determination of the next event for a
given particle is impractical in our large simulations because $O(N)$ work is required {\it per particle}
to examine all possible collision partners, where $N$ is the total number of particles.

Fortunately this work can be reduced to a constant independent of $N$
by dividing the bubble into a number of cells \cite{Rapaport,erp77}.  
By taking an edge length
that is larger than the sphere diameter, $\sigma$, it is obvious that
collisions can only occur between particles in the same and adjacent cells.
Since we want a relatively small number of particles in each cell
{\it and} we want the number of cells to be comparable to the number
of particles in the simulation
the size of the cells must be reduced as the bubble collapses.
We use a straightforward subdivision procedure to accomplish this task.
Initially, the bubble is subdivided
into approximately $8N$ square, identical cells.  Every time the bubble 
declines by a factor of two in diameter, cell size is reduced by a factor
of two (keeping in mind that we must stop the procedure once the cell diameters
reach the particle diameter).  Note that we do not need to recompute which particle belongs to
which cell after each collision.  Instead, we 
introduce a cell crossing event and update
a particle's cell location only when the corresponding cell crossing event is processed (cf. \cite{Rapaport,erp77}).

Also note that when a collision occurs, only particles in the
immediate neighborhood of the cell need to be updated.  For this reason
a `personal' time is stored for each particle, representing the time when the
particle was last updated.  The entire system configuration only needs to be updated
when properties (such as density, temperature etc.) are evaluated. \cite{Rapaport,erp77}

\subsection{Event Calendar}
Because we require information on when particle collisions,
hard wall collisions and cell crossings occur some sort of {\it event calendar}
is needed.  This calendar will store many future events.  As
collisions and cell crossings occur, newly predicted collisions
and cell crossings must be added to the calendar and events 
that are no longer relevant must be removed.  \cite{Rapaport}

Specifically, whenever particles collide their velocities are changed.
This implies that future events involving these particles
are no longer valid and should be removed from the calendar.  
It also implies that new cell crossings
and collision events need to be calculated and added to the calendar.  
(Fortunately, the only particle collisions that need to be considered are those 
involving the current cell and its neighbors.)
On the other hand, when a particle crosses cell boundaries 
previous particle collisions remain valid but the newly adjacent cells
contain potential collision targets that must be examined.  New cell crossing
events must also be checked.
As pointed out in \cite{Rapaport},
there is no way in which a collision can be missed
provided all these details are taken care of
correctly.

Of course, it is essential that the calendar can be managed efficiently
both in terms of memory and CPU usage.
To meet this requirement, we utilize the binary tree data structure 
described in \cite{rap80,Rapaport}.  Here, each scheduled event
is represented by a node in the tree.  The information
contained within the node identifies the
time at which the event is scheduled and the event details.
Calendar events are added or deleted by adding or deleting
the corresponding nodes from the tree.  To facilitate traversing the
tree, three pointers are used to link event nodes.
These point to the left and right descendents of the node and to the node's parent.
The ordering is carried out so that left hand descendents of a particular node
are events scheduled to occur before the event at the current node, while right hand descendents
correspond to events that occur after it.  Finally, the rapid deletion of event nodes
is supported by linking event nodes into two circular lists.  See \cite{Rapaport}
for details.

It is interesting that estimates of the theoretical performance of 
the tree structure are possible in a number of instances \cite{knu73,Rapaport}.  For example, if
a tree is constructed from a series of events that are randomly distributed,
the average number of nodal tests to insert a new node into the tree is $2\log N$.
Also, the average number of cycles to delete a randomly selected
node is a constant independent of $N$.  It is noteworthy that measurements 
have been performed to confirm these results
in actual MD simulations \cite{rap80}.
Our sonoluminescence simulations spend most of the CPU time on
compressing the bubble from its maximum radius to the ambient radius.  Since the
bubble is fairly uniform in this regime,
the assumption of a random distribution of events seems plausible and we expect
that this type of estimate on theoretical performance should hold.
(On the other hand, near the short-lived hot spot the behavior is far from
equilibrium and this assumption on randomness may be invalid.)
A detailed study of the theoretical performance of the tree structure
will be the focus of subsequent work.

\subsection{Average Properties}
We need to evaluate spatially dependent average properties
of the gas at various times.  To minimize
statistical fluctuations, we assume that the results
are radially symmetric and average over shells
that are 1/40th of the bubble radius.
We calculate dimensionless values for
density, temperature, velocity and average charge as follows:
\begin{itemize}
\item The dimensionless density is given by the density divided by the average ambient density.
\item The dimensionless velocity is given by the velocity divided by the ambient speed of sound $\sqrt{\gamma k T_0/m}$
where $\gamma$ is the ratio of heat capacities and $m$ is the mass of a single particle.
\item The dimensionless temperature is given by the temperature divided by the ambient temperature, $T_0$.
Specifically,
\[
T = \frac{m}{3Nk T_0} \sum_i^N (v_i^2 - v_n^2)
\]
where the summation is over all $N$ particles in the shell, $v_i$ is the speed of
the $i^{th}$ particle and $v_n$ is the normal speed of the gas in the shell.
\item Ionization is simply the average charge per particle.
\item In each case we plot properties as a function of a dimensionless bubble radius, $r$, 
which equals the physical radius $R(t)$ divided by a constant approximating
the atomic diameter.  For Helium this constant is chosen to be 2.18\AA~(see 
Table~I).  For Argon and Xenon these constants are chosen to be 4.11\AA~and 
5.65\AA, respectively.  (These latter two choices represent average VSS model values 
at 273K and also approximate the values
given in Table~I).
\end{itemize}
See \cite{Rapaport} for further details on calculating equilibrium
and transport properties for hard sphere models.

\section{Simulation Results} \label{sec:results}
In this section we simulate the collapse of a
sonoluminescing bubble from its maximum radius to its
hot spot.  Our focus is on how 
boundary conditions affect the interior dynamics of the collapse.
Results for Helium, Argon and Xenon
are presented. 

The section begins with a study of the collapse of million
particle bubbles
and concludes by addressing how simulations vary according to ambient
bubble size.

\subsection{Helium Bubbles with Specular BCs}
We first consider evolving a bubble of one million 
helium atoms using specular boundary conditions.

\begin{figure}
 \vspace*{2.75in}
 \includegraphics{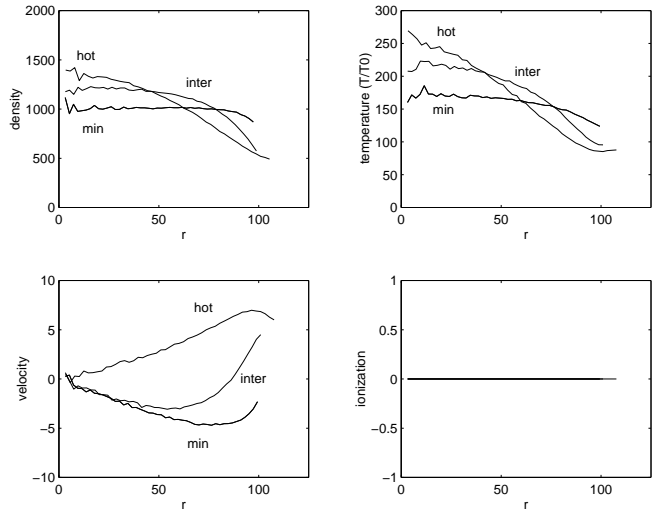}
 \caption{Helium bubble with specular BCs, constant diameter particles and
no ionization.  Here, $R_{min} = 99.4$, $R_{inter} = 101.0$ and $R_{hot} = 107.7$} \label{fig:noions}
\end{figure}

With the {\it constant diameter model} and {\it no ionization} 
the temperature and density increase uniformly as the bubble
collapses to the minimum radius.  
After the minimum radius is attained, the temperature becomes
hotter towards the center of the bubble and cools at the
expanding outer boundary of the bubble, with a peak temperature
of about 80000K reached at the center.  
(At these temperatures, it is clear that ionization events will occur so
the remainder of our simulations consider ionization.)
See Figure~\ref{fig:noions}
for plots of the density, temperature and velocity as a function of distance from the 
center of the bubble at various bubble radii.

\begin{figure}
 \vspace*{2.75in}
 \includegraphics{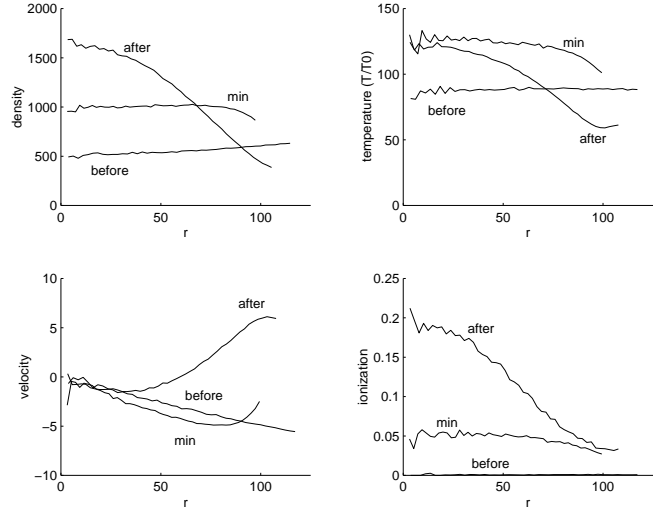}
 \caption{Helium bubble with specular BCs, constant diameter particles and
ionization.  Here, $R_{before} = 117.2$, $R_{min} = 99.4$ and $R_{after} = 107.7$} \label{fig:ions}
\end{figure}

With the {\it constant diameter model} and {\it ionization} 
we again find that the temperature and density 
increase uniformly as the bubble
collapses to the minimum radius.  
However, after the minimum radius is attained, ionization
causes the temperature to cool
across the entire bubble rather than just at the
boundary of the bubble (although cooling occurs most rapidly
at the bubble boundary).  A peak temperature
of about 40000K is attained at the minimum radius.
It is particularly noteworthy that recorded properties 
are nearly constant throughout the bubble when the peak temperature
occurs --- See Figure~\ref{fig:ions}.

\begin{figure}
 \vspace*{2.75in}
 \includegraphics{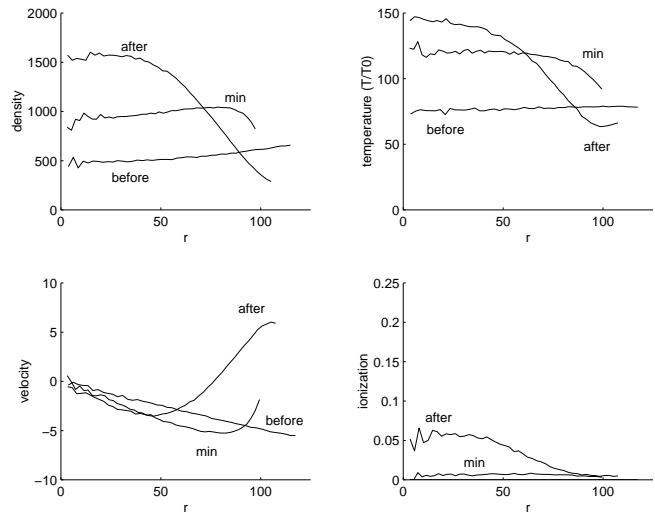}
 \caption{Helium bubble with specular BCs, VSS diameter particles and
ionization.  Here, $R_{before} = 117.4$, $R_{min} = 99.4$ and $R_{after} = 107.5$} \label{fig:he_specular_vss}
\end{figure}

Changing to the {\it VSS diameter model} 
gives very similar results, except now ionization occurs
less frequently because the effective size of the
particles is smaller.  Because less ionization occurs,
the temperature continues to increase
for a short while after the minimum bubble radius
leading to a peak temperature
of about 45000K.  See Figure~\ref{fig:he_specular_vss}
for details.

\subsection{Helium Bubbles with Heat Bath BCs}
Our next set of simulations evolve a bubble of one million 
helium atoms using heat bath boundary conditions and ionization.

\begin{figure}
 \vspace*{2.75in}
 \includegraphics{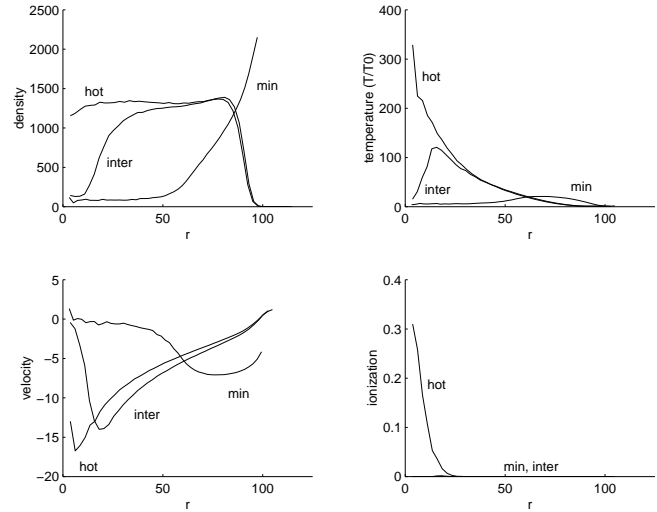}
 \caption{Helium bubble with heat bath BCs, constant diameter particles and
ionization.  
Here, $R_{min} = 99.4$, $R_{inter} = 114.8$ and $R_{hot} =117.2$} \label{fig:he_const_bath}
\end{figure}

With the {\it constant diameter model}
the density increases dramatically at the edge of the
bubble as the minimum radius is attained.  Temperature
and velocity are also much more profiled than for 
specular boundary conditions, with peaks occurring about
25 percent of the way from the boundary of the bubble to
the center.  No ionization has occurred at the minimum radius.
For a short time after the minimum radius\footnote{Note that a vacuum forms
at the bubble wall after the minimum radius occurs.  This is simply an
artifact of using Equation~\ref{eq:rayleigh} as the forcing equation 
for $R(t)$.},
the peak temperature of the bubble
continues to increase (to a maximum of 95000K), 
and temperature and density profiles
become even more pronounced --- See Figure~\ref{fig:he_const_bath}.
At first sight, it is counterintuitive that
heat bath boundaries create conditions
whereby the cooling from the boundary
leads to greater energy focusing and higher peak temperatures.
Perhaps cooling lowers the speed of sound and enhances the nonlinear
response to the high speed $\dot{R}$ of collapse.

\begin{figure}
 \vspace*{2.75in}
 \includegraphics{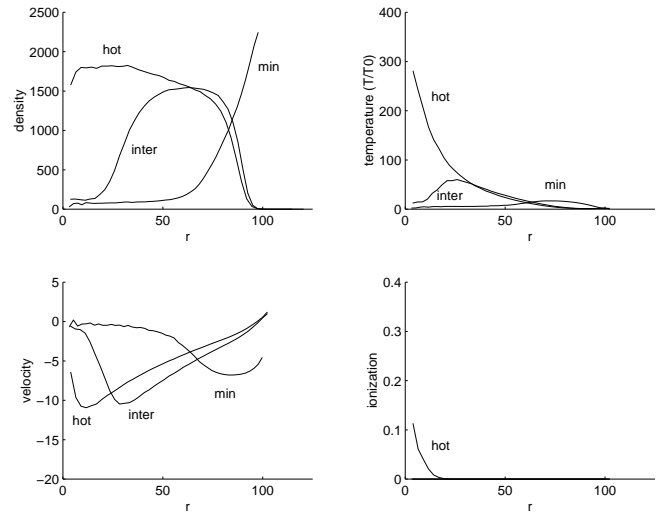}
 \caption{Helium bubble with heat bath BCs, VSS diameter particles and
ionization.  Here, $R_{min} = 99.4$, $R_{inter} = 117.2$ and $R_{hot} =123$} \label{fig:he_bath_vss}
\end{figure}

Changing to the {\it VSS diameter model} 
gives very similar results, except now ionization occurs
less frequently because the effective size of the
particles is smaller.  
See Figure~\ref{fig:he_bath_vss} for details.

\subsection{Argon and Xenon Bubbles with Specular BCs}
Our next set of simulations evolve million particle Argon
and Xenon bubbles using specular boundary conditions and ionization.

\begin{figure}
 \vspace*{2.75in}
 \includegraphics{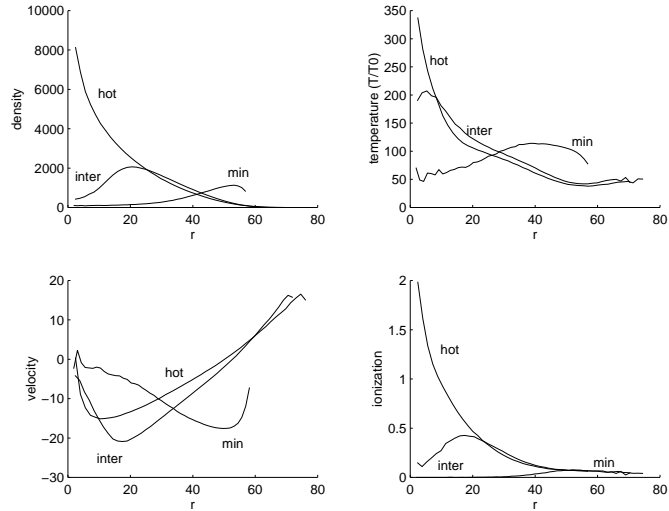}
 \caption{Argon bubble with specular BCs, VSS diameter particles and
ionization.  Here, $R_{min} = 58.2$, $R_{inter} = 72.0$ and $R_{hot} =76.2$} \label{fig:ar_specular_vss}
\end{figure}

We start by considering an Argon bubble with the {\it VSS diameter model}.
Because the speed of sound is slower in Argon than in Helium we expect
Argon simulations to exhibit much sharper profiles than
Helium.  This is indeed the case.  Moreover,
our simulation results are surprisingly similar to those for
Helium with {\it heat bath boundaries}:
Density increases at the edge of the
bubble as the minimum radius is attained.  Temperature
and velocity are sharply profiled, with peaks occurring 
closer to the boundary of the bubble than to its center.
Also, for a short time after the minimum radius,
the peak temperature of the bubble
continues to increase rapidly (to a maximum of 100000K), 
and temperature and density profiles
become even more pronounced --- See Figure~\ref{fig:ar_specular_vss}.

\begin{figure}
 \vspace*{2.75in}
 \includegraphics{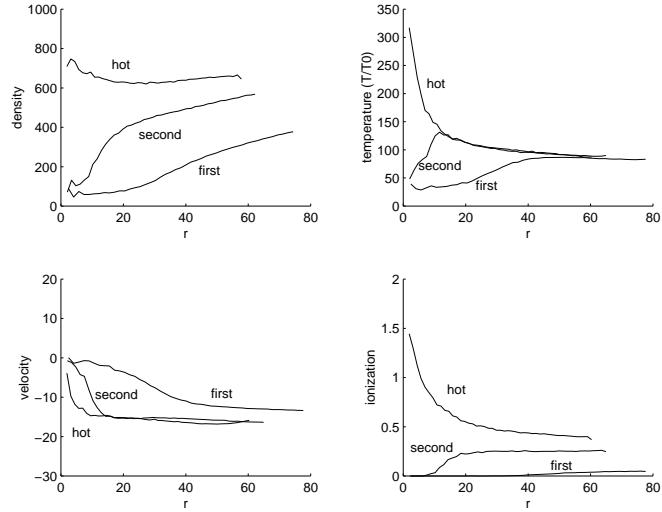}
 \caption{Argon bubble with specular BCs, constant diameter particles and
ionization.  Here, $R_{first} = 77.6$, $R_{second} = 64.9$ and $R_{hot} =60.3$} \label{fig:ar_specular_const}
\end{figure}

Constant diameter hard sphere simulations of Argon are also possible.
These simulations are 
unique\footnote{This behavior may be related to the consistency of the minimum
bubble radius and the hard sphere radius.  See the
case of Xenon below.}
in that the hot spot occurs before the minimum radius value of 58.2 ---
See Figure~\ref{fig:ar_specular_const}.  
As expected, this simulation
gives sharper profiles than the corresponding model for Helium.
However, since the minimum radius is close to the minimum radius allowed
by the packing of the hard spheres
the results are much more uniform than those derived using the VSS model
for Argon.  Also note that as a result of the collapse of the bubble, energy stored
at the maximum radius is converted into heating, ionization and kinetic
energy of the local center of mass.  From Figure~\ref{fig:ar_specular_const}
one can estimate these quantities.  The average temperature of the atoms is 30000K
which is a thermal energy of about 3.75 eV/atom.  As half the atoms
are ionized, the ionization energy is about 8 eV/atom.  Since electrons have about
the same thermal energy as the ions, their energy is about 2 eV/atom.  Taken together,
these channels add to about 21 eV/atom which is less than the 25 eV/atom available
in the initial state but the difference is within the accuracy of the energy estimates. For Helium
at the hot spot, the energy of the hard sphere (plus ionization) is substantially
less than the energy stored at $R_m$.  This can be attributed to the fact that
$a_{hs}/a \approx 0.60$.  For Argon, almost all the stored energy ends up in the
hard sphere gas since $a_{hs}$ is much closer to $a$; $a_{hs}/a \approx 0.91$.
In both cases, inclusion of a self-consistent boundary condition at the wall will
account for any further energy discrepancies.  Of course, in a physical system,
energy will diminish due to acoustic radiation 
and thermal losses through the boundary of the bubble and we expect this to constitute
a strong effect.

\begin{figure}
 \vspace*{2.75in}
 \includegraphics{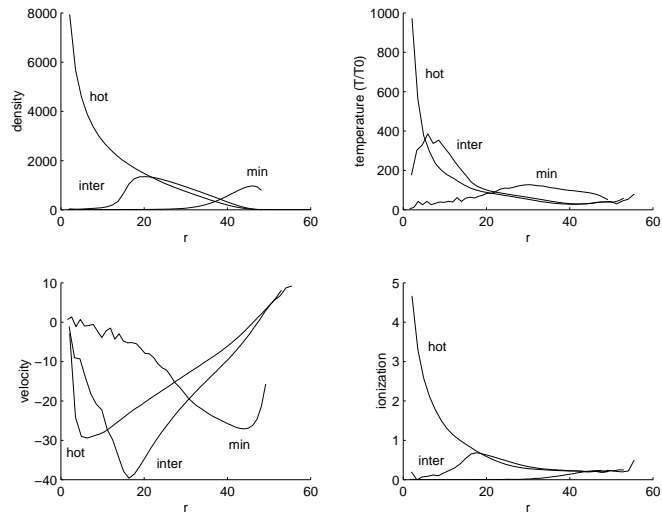}
 \caption{Xenon bubble with specular BCs, VSS diameter particles and
ionization.  Here, $R_{min} = 49.2$, $R_{inter} = 62.2$ and $R_{hot} =66.7$} \label{fig:xe_specular_vss}
\end{figure}

Simulations for Xenon bubbles with the {\it VSS diameter model}
were also carried out.  
Because the speed of sound is slower in Xenon than in Argon we expect
Xenon simulations to exhibit even sharper profiles than Argon.  
Indeed, this is the case and temperatures of up to
300000K were obtained despite the occurrence of 
multiple ionization (exceeding 4 per particle at the center) ---
See Figure~\ref{fig:xe_specular_vss}.

A proviso for the Xenon data is that these calculations bog 
down before the minimum radius is attained
when the constant diameter model is used, whereas the Helium data is hardly affected 
by this modification.  The explanation lies in the consistency of the 
minimum bubble radius $(\sim a)$ and the hard sphere radius for Xenon.  Specifically,
the minimum radius of the bubble wall is less than the minimum
radius allowed by the packing of hard spheres.
On the other hand, Xenon simulations carried out using the
VSS model are relatively insensitive to changes in $a$.  For example,
increasing $a$ by $30\%$ 
changes the peak temperature by about $35\%$, and leaves the qualitative features invariant.
Note that in this case, $a_{hs}<a$ as with Helium simulations.

\subsection{Argon and Xenon Bubbles with Heat Bath BCs}
We now consider the evolution of million particle Argon
and Xenon bubbles using Heat Bath boundary conditions and ionization.

\begin{figure}
 \vspace*{2.75in}
 \includegraphics{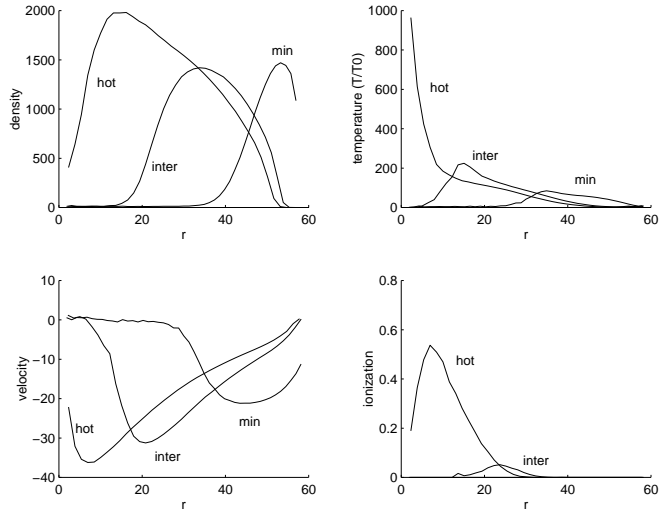}
 \caption{Argon bubble with heat bath BCs, VSS diameter particles and
ionization.  Here, $R_{min} = 58.2$, $R_{inter} = 68.1$ and $R_{hot} =73.2$} \label{fig:ar_bath_vss}
\end{figure}

Applying the {\it VSS diameter model} to an Argon bubble 
gives results that have the same qualitative features as the corresponding
Helium simulation, except that all properties are
much more sharply profiled.
Indeed, temperatures of up to
300000K were obtained in this simulation
showing (once again) that heat bath boundaries create conditions
whereby the cooling from the boundary
leads to greater energy focusing and higher peak temperatures
than specular boundary conditions.
See Figure~\ref{fig:ar_bath_vss} for details.

\begin{figure}
 \vspace*{2.75in}
 \includegraphics{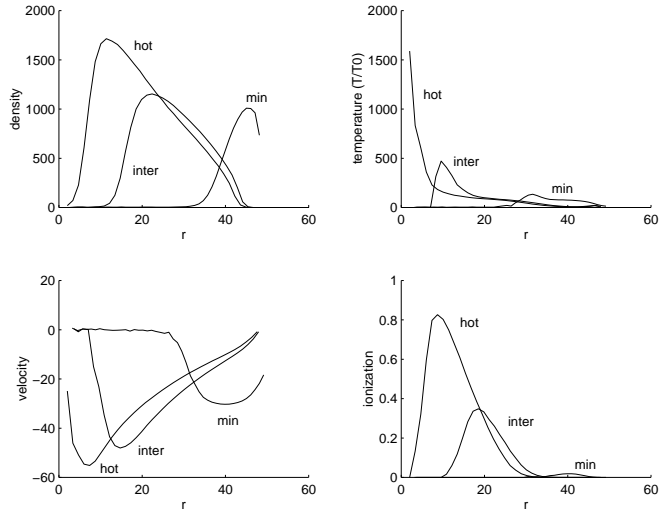}
 \caption{Xenon bubble with heat bath BCs, VSS diameter particles and
ionization.  Here, $R_{min} = 49.2$, $R_{inter} = 60.8$ and $R_{hot} =63.7$} \label{fig:xe_bath_vss}
\end{figure}

Simulations for Xenon bubbles with the {\it VSS diameter model}
were also carried out.  
Because the speed of sound is slower in Xenon than in Argon,
Xenon simulations exhibit even sharper profiles than
Argon.  Indeed, temperatures of up to
500000K were obtained.
See Figure~\ref{fig:xe_bath_vss} for details.

As discussed in the previous section, the constant diameter model for
Xenon is not able to compute down to the minimum radius since
that radius is smaller than the minimum packing radius of the hard sphere gas.
On the other hand, Argon calculations are possible.
These simulations are qualitatively similar to the constant diameter model for
Helium, but
produce a more highly ionized gas (reaching an average charge of +6 per particle near the center)
and much higher temperatures (up to 1.5 million K) than any other simulations
that were considered.  Because such extreme values arise it seems likely
the constant diameter model for Argon also experiences a significant consistency problem
near the minimum radius.

\subsection{Flashwidths}
An important experimental measurement for SL bubbles is the flash width, i.e. the duration of the light emission,
because this constrains the possible light emission mechanisms and thus provides a point of
validation for any proposed model or theory.
Since our simulations do not include the fundamental atomic excitation or charge acceleration effects responsible
for radiation, the current model does not directly yield a flashwidth.

However, an estimated flash width can be obtained from the computed temperature as a function of time.  
If we assume that whatever process is responsible for the light emission is strongly dependent on the
current temperature, and that it does not appreciably alter the gross gas dynamics, the flashwidth 
at a particular color is simply the length of time which the temperature exceeds the appropriate turn-on
threshold. In this case, the peak temperature as a function of time is our key diagnostic quantity.

\begin{figure}
 \vspace*{2.in}
 \includegraphics{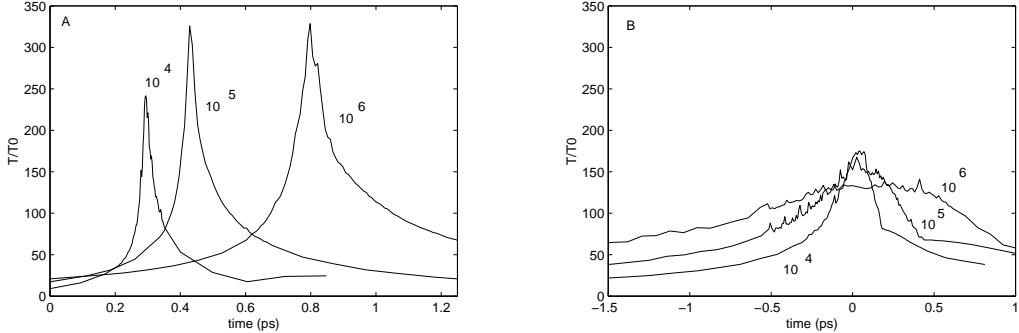}
 \caption{Helium peak bubble temperature vs. time with $time=0$ marking the time corresponding
to the minimum radius.  (A) Heat Bath boundary conditions.
(B) Specular Boundary conditions.} \label{fig:flashwidths}
\end{figure}

Our simulations (Figure~\ref{fig:flashwidths}) show that emission from an adiabatic compression
lacks a strong, sharp temperate spike in time, and thus the associated flash from this
model would be
longer and would be comprised of lower energy photons. 
In contrast, the heat bath boundary conditions yield sharp transient spike in temperature, and
thus this model predicts a much shorter flash which is comprised of higher energy photons. 
In both cases it appears that
the width of the spike roughly doubles as the number of particles in the bubble
increases by factors of 10, from $N = 10^4$ to $N = 10^6$. Since each factor of $8$ in particle number corresponds to
a doubling of the ambient bubble radius $R_0$, this amounts to essentially a predicted linear scaling
between flash width and ambient bubble radius.

In both plots, the curves for $N = 10^5$ and $N = 10^4$ particles
were derived by averaging over 10 and 20 simulations respectively in order to keep
statistical fluctuations to an acceptable level. The  $N = 10^6$ simulation required
just a single simulation for robust statistics.

\section{Summary and Future Work}
Sonoluminescence is well suited to investigation by Molecular Dynamics
because the range of densities and time scales is large, yet the number of particles 
involved is relatively small. Because the phenomena still poses
experimentally difficult, unsolved questions regarding its mechanism and ultimate energy focusing energy potential, 
we feel it is an excellent subject for much more detailed MD investigations
than the initial effort we have presented here.

In this paper, we introduced a simple model for the interior dynamics of single noble gas bubble
sonoluminescence, 
as a hard sphere gas driven by a spherical piston controlled be the  Rayleigh-Plesset equation.
Energy losses due to ionization were also accounted for.
We considered both constant and variable radius hard sphere models, and these lead to
quantitatively similar results. Fast, tree-based algorithms allowed us to evolve million particle systems
through the entire collapse process.
Our calculations indicate that extreme energy focusing occurs within the bubble which in some cases is driven
by a shock-like compression in the gas.  Peak temperatures range
from 40,000 K for He to 500,000 K for Xe. These are accompanied by high levels of ionization
during the final collapse, and formation of a transient, high density plasma state seems quite likely.

The imposition of a thermal boundary condition at the wall of the bubble
leads to greatly increased energy focusing and non-uniformity within 
a collapsing bubble. In any case, the predicted flash width
scales roughly linearly with the ambient bubble radius.

There are a variety of interesting directions for future research in this problem.
For example, our simulations simply treat
the bubble wall as a piston moving in with a prescribed velocity.
A natural improvement would be to couple the
internal molecular dynamics to the wall velocity
to obtain a self-consistent bubble motion and internal dynamics. 
This could be done by coupling to Euler or Navier Stokes models for the
surrounding fluid.
This may be particularly important for accurately computing the dynamics
through the point of minimum radius.
In our present model there may be over compression of the gas in some simulations
as the minimum radius is approached,
since the prescribed piston motion does not respond to the rapid increase in the internal gas
pressure. Conversely, when the piston retracts after this point, a nonphysical gap often develops
between the bubble boundary and the outer extent of the gas, which may under-compress the gas.
 
Another important area for future research is adding
in water vapor into the bubble interior. This provides a potentially important 
cooling mechanism, which may strongly modulate the light emission and 
energy focusing, and may explain the strong ambient temperature dependence of the emitted light intensity. 
We have done preliminary investigations of this,
and these show that water evaporation from the bubble surface into
the interior must be included to avoid rapid expulsion of the water vapor.
It is also possible that the water could be
directly involved in the light emission, when it is properly included in the model.

Other bubble collapse geometries could also be considered, and these may have
different energy focusing characteristics. For example, one could consider
a nonspherical collapse, hemi-spherical bubbles collapsing on a solid surface, or consider collapse geometries
appropriate for bubble jetting scenarios. Similarly, one could see if special collapse
profiles can be used to reach much higher internal temperatures, and otherwise
explore the extremes of the energy focusing potential. Perhaps a mode could even be
found in which small amounts of deuterium-deuterium fusion could be induced, 
assuming there is deuterium gas in the bubble as well.

Including additional atomic physics such as atomic excitation, rotational
and vibrational degrees of freedom (needed for non-noble gases or water vapor)
and electron-ion recombination would all allow for more accurate energy accounting,
and may also be directly related to light emission mechanisms.

Another major direction would be to include electric field effects into the
the simulation.  Algorithms for such models must treat long range electrostatic interactions 
to avoid incurring serious errors.
They must also be able to evaluate long range forces efficiently since calculating
interactions pairwise becomes expensive for more than a few thousand particles.
For these reasons, multipole methods are particularly attractive---
they use a hierarchy of spatial subdivisions and a multipole expansion
to evaluate interactions with little more than linear effort in the number of particles.  See
\cite{Greengard1,Greengard2} for details and see
also \cite{Rapaport} for further references on methods for evaluating long range forces.
This would potentially allow direct simulation of the ions and electrons produced 
as well, which,  along with including atomic excitation, would allow direct simulation
of the light emitting processes. With these effects included, an extremely detailed
picture of the SL phenomena could be laid out.
 
Of course, larger scale, parallel simulations are essential 
to actually achieve direct comparisons with present SL experiments. Because
the simple hard sphere interactions are quite local, the system should be amenable to
parallelization. 
We expect that the cost (in collision count) for a hard sphere MD simulation scales
roughly like $N^{4/3}$, where $N$ is the number of particles, since the collapse
time from the ambient to the minimum radius scales linearly with $R_0$ (and so we
conjecture that the collision rate also increases roughly linearly with the $R_0$).
Thus simulations using 
one hundred times as many particles (i.e. $N=10^8$) 
would require $500$ times as much computer time assuming near optimality
in the algorithm. This is somewhat beyond the 
range of a single supercomputer CPU, but would become quite practical on a 100 
node system of workstation-grade CPUs.

Finally, it would be of great interest to investigate where less costly 
continuum models and Monte Carlo simulations are appropriate 
for studying sonoluminescence and to develop techniques for coupling these methods to
detailed molecular dynamics simulations near the light emitting hot spot, in order to produce
more complete models with greater predictive validity.

There is also a great deal to explore experimentally.
One example relevant to our study is that it would be useful to measure
flashwidth as a function of ambient bubble radius (or, in practice, 
intensity and frequency of the driving sound field), for comparison with 
the scaling predictions of MD and other models.
 
\section{Acknowledgments}
We thank D.B. Hash, A.L. Garcia and P.H. Roberts for valuable discussions.

\bibliographystyle{plain}
\bibliography{biblio}
\end{document}